\documentclass{sigchi} 

\usepackage[bottom]{footmisc}
\usepackage{ifpdf}
\usepackage{cite}
\usepackage{algorithmic}
\usepackage{array}
\usepackage{bm}
\usepackage{mdwmath}
\usepackage{mdwtab}
\usepackage{eqparbox}
\usepackage{subfigure}
\usepackage{multirow}
\usepackage[lined, algonl, boxed]{algorithm2e}
\usepackage{booktabs}

\newcommand{\paragraphX}[1]{\vskip 0pt \noindent \textbf{#1} \hskip .05in}

\newcommand{\note}[1]{\fxnote{\hl{#1}}}

\usepackage[final,inline,nomargin]{fixme}      

\usepackage{color}
\usepackage{soul}          

\usepackage{balance}  
\usepackage{graphics} 
\usepackage{times}    
\usepackage{url}      

\makeatletter
\def\url@leostyle{%
  \@ifundefined{selectfont}{\def\UrlFont{\sf}}{\def\UrlFont{\small\bf\ttfamily}}}
\makeatother
\def\pprw{8.5in}
\def\pprh{11in}

\usepackage[pdftex]{hyperref}
\hypersetup{
pdftitle={SIGCHI Conference Proceedings Format},
pdfauthor={LaTeX},
pdfkeywords={SIGCHI, proceedings, archival format},
bookmarksnumbered,
pdfstartview={FitH},
colorlinks,
citecolor=black,
filecolor=black,
linkcolor=black,
urlcolor=black,
breaklinks=true,
}

\begin{document}

\title{Towards Making Random Passwords Memorable: Leveraging Users' Cognitive Ability Through Multiple Cues}

\numberofauthors{3}
\author{
  \alignauthor Mahdi Nasrullah Al-Ameen\\
    \affaddr{Department of CSE}\\
    \affaddr{The University of Texas at Arlington}\\
     \affaddr{Arlington, TX, USA}\\      
    \email{mahdi.al-ameen@mavs.uta.edu}\\
   \alignauthor Matthew Wright\\
     \affaddr{Department of CSE}\\
     \affaddr{The University of Texas at Arlington}\\
     \affaddr{Arlington, TX, USA}\\      
     \email{mwright@uta.edu}   
  \alignauthor Shannon Scielzo\\
      \affaddr{Department of Psychology}\\
      \affaddr{The University of Texas at Arlington}\\
      \affaddr{Arlington, TX, USA}\\       
      \email{scielzo@uta.edu}
}

\maketitle

\begin{abstract}
Given the choice, users produce passwords reflecting common strategies and patterns that ease recall but offer uncertain and often weak security. System-assigned passwords provide measurable security but suffer from poor memorability. To address this usability-security tension, we argue that systems should assign random passwords but also help with memorization and recall. We investigate the feasibility of this approach with {\em CuedR}, a novel {\em cued-recognition} authentication scheme that provides users with multiple cues (visual, verbal, and spatial) and lets them choose the cues that best fit their learning process for later recognition of system-assigned keywords. In our lab study, all $37$ of our participants could log in within three attempts one week after registration (mean login time: $38.0$ seconds). A pilot study on using multiple CuedR passwords also showed $100\%$ recall within three attempts. Based on our results, we suggest appropriate applications for CuedR, such as financial and e-commerce accounts.

\end{abstract}


 {Authentication} 

\section{Introduction} \label{intro}

In most systems, users are tasked with creating a password that should be both secure and memorable. Users, however, typically lack information about what is secure in the face of modern cracking and attacks tools, as well as how to construct memorable strings, memorize them quickly, and accurately recall them later. Faced with this challenge, users often create passwords that may seem secure and memorable but fail on one or both counts. Failure to understand security requirements leads to guessable passwords, while memorability issues lead not only to inconvenience, but also to password reset systems that are often abused by hackers~\cite{palin_hack,no_sec}.

We argue that the burden of password creation should be borne by the system, rather than the user. With system-assigned passwords, the user does not have to guess whether a password is secure, and the system can ensure that all passwords offer the desired level of security. Additionally, while password reuse could pose a serious security threat~\cite{pwreuse14}, using system-assigned passwords ensures that users do not reuse a password (or modification thereof) already used on another account.

Making system-assigned passwords memorable, however, has proved challenging. Different variants of system-assigned passwords have been proposed~\cite{passphrase,PTP,text_recog,interference3}, but none of them provides sufficient memorability. We postulate that new authentication systems should more effectively make use of humans' cognitive strengths and accommodate users with different learning styles. To this end, we draw upon several prominent theories of memory to design {\em CuedR}, a novel authentication scheme that offers visual, verbal, and spatial cues to help users recognize system-assigned keywords.

\subsection{Contributions} 
{\em Memorability.} In CuedR, the system assigns users six keywords, each from a distinct portfolio (e.g., animals, fruits, or vehicles) of $26$ keywords. Both at registration and at login, users are provided with an image of the keyword (a visual cue); a number and a phrase associated with the keyword (verbal cues); and the fixed position of all of the elements on the page (spatial cues). Users with different learning styles can focus on the cues that help them best remember the keyword. Moreover, the cues facilitate an elaborative encoding that helps to transfer the keywords from the working memory to long term memory at registration~\cite{long_mem68}, helping users recognize their keywords when logging in later.

In our single-password study, all $37$ participants remembered their CuedR password after one week of registration. We note that no other system-assigned password scheme has reported $100$\% memorability to our knowledge, even schemes offering only PIN-replacement security levels (e.g. $13$ bits of entropy). Despite high login times ($38.0$ seconds on average), participants reported high levels of satisfaction with the scheme and $84$\% preferred to use it in real life as a replacement to traditional textual passwords. 

{\em Security.} By using system-assigned random passwords, the effective entropy of the passwords is equal to the theoretical entropy, which is set to $28$ bits in our studies. Additionally, CuedR provides \textit{variant response} during login, which is known to be an important feature to gain resilience against observation attacks (e.g., shoulder surfing, keystroke loggers)~\cite{survey}.

\section{Related Work}\label{background} 

In this section, we give a brief overview of notable textual and graphical password schemes, in which we highlight why existing schemes are insufficient. 

\subsection{Textual Password Schemes}
\vspace{0.1cm}
\textbf{User-chosen passwords.} Traditional user-chosen textual passwords are fraught with security problems and are especially prone to password reuse and predictable patterns~\cite{pwreuse14,pwpattern1}. Das et al.~\cite{pwreuse14} found that $43$\% of users use the identical password in multiple sites, while $30$\% of non-identical passwords could be cracked in less than $100$ attempts. Shay et al.~\cite{pwpattern1} report that password restriction policies do not necessarily lead to more secure passwords but can adversely affect memorability. 

\textbf{System-assigned passwords.} System-assigned random textual password schemes are more secure but fail to provide sufficient memorability, even when natural-language words are used~\cite{passphrase,text_recog}. Wright et al.~\cite{text_recog} compared the usability of three different system-assigned textual password schemes: Word Recall, Word Recognition, and Letter Recall. None of these schemes had sufficient memorability rates. Forget et al.~\cite{PTP,forget_thesis} proposed the Persuasive Text Passwords (PTP) scheme as a hybrid between user-selected and system-assigned passwords, in which the user first creates a password and PTP improves its security by placing randomly-chosen characters at random positions in the password. Unfortunately, the memorability for PTP is just $25$\% when two random characters are inserted~\cite{forget_thesis}.

\begin{figure}[t]
\centering
\includegraphics[width=89mm,height=62mm]{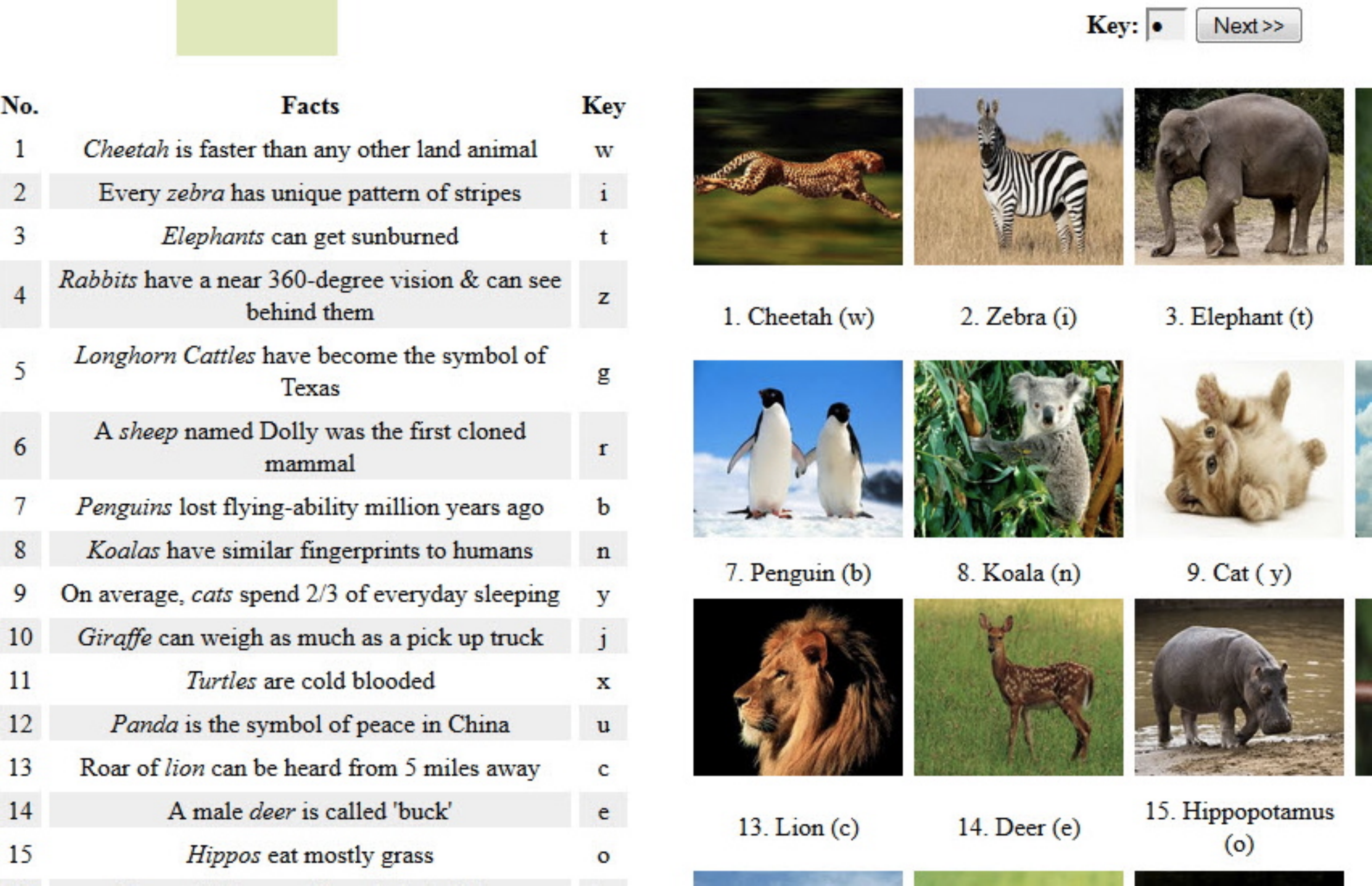}
\caption{A partial screen shot during login. The facts corresponding to each keyword appear on the left side of the screen. The key is shown in parenthesis next to each keyword and also in the rightmost column of the table}
\label{fig:portfolio}
\end{figure}

\subsection{Graphical Password Schemes}

Graphical password schemes can be divided into three categories~\cite{survey}, based on the kind of memory leveraged by the systems: i) Drawmetric (recall-based), ii) Locimetric (cued-recall-based), and iii) Cognometric (recognition-based). \textit{Passfaces}~\cite{passface}, a cognometric graphical password scheme, is commercially available and deployed by a number of organizations, including banks and government agencies.\footnote{\url{http://www.realuser.com/} shows testimonials about Passfaces from customers.}

\textbf{Drawmetric.} The user is asked to reproduce a drawing in this category of graphical passwords. In \textit{Draw-a-Secret (DAS)}~\cite{das}, a user draws on top of a grid, and the password is represented as the sequence of grid squares. Nali and Thorpe~\cite{das2} have shown that users choose predictable patterns in DAS. \textit{BDAS}~\cite{bdas} intends to reduce the amount of symmetry in the user's drawing by adding background images, but this may introduce other predictable behaviors such as targeting similar areas of the images or image-specific patterns~\cite{survey}. DAS and BDAS have recall rates of no higher than $80\%$. 

\textbf{Locimetric.} The password schemes in this category, including {\em Passpoints} and {\em Cued Click-Points (CCP)}, present users with an image and have users select points on the image as their password. Dirik et al.~\cite{passpoint4} developed a model that can predict 70-80\% of users' click positions in Passpoints. To address this issue, Chiasson et al. proposed \textit{Persuasive Cued Click-Points (PCCP)}~\cite{pccp_pilot,pccp2}, in which a randomly-positioned viewport is shown on top of the image during password creation, and users select their click-point within this viewport. The memorability for PCCP was found to be 83-94\%. In a follow-up study, Chiasson et al.~\cite{click_pattern} found predictability in users' click points and indicate that predictability is still a security concern for PCCP.

\textbf{Cognometric.} In this recognition-based category of graphical passwords, the user is asked to recognize and identify their password images from a set of distractor images. \textit{Passfaces}~\cite{passface} is a commercial cognometric system in which users select one face among a panel of nine distractor faces and repeat this over several panels. Davis et al.~\cite{story} have found that users select predictable faces, biased by race, gender, and attractiveness of faces. As a result, the commercial Passfaces~\cite{passface} product now assigns a random set of faces instead of allowing users to choose. However, Everitt et al.~\cite{interference3} show that users have difficulty in remembering system-assigned Passfaces. Hlywa et al.~\cite{image_type} found no significant difference in memorability between cognometric schemes  providing either face images or object images (entropy: $28$ bits), while the mean login time was $31$ seconds for object recognition and $41$ seconds for face recognition. 
\note{mentioned about login time of prior cognometric schemes providing same entropy as CuedR (R2)} 

In sum, schemes with lower risk of predictability also show lower recall rates. Password managers~\cite{pw_manager} fail to provide a suitable solution in this case, as it suffers from usability (in implementation) and security (e.g., single point of failure) problems. Indeed, two recent papers extensively examine security problems in a range of password managers~\cite{pm1,pm2}. 

Thus, despite a large body of research, it remains a critical challenge to build an authentication system that offers both high memorability and guessing resilience.

\section{C\lowercase{ued}R: System Design}

\begin{figure}[t]
\centering
\includegraphics[width=87mm, height=45mm]{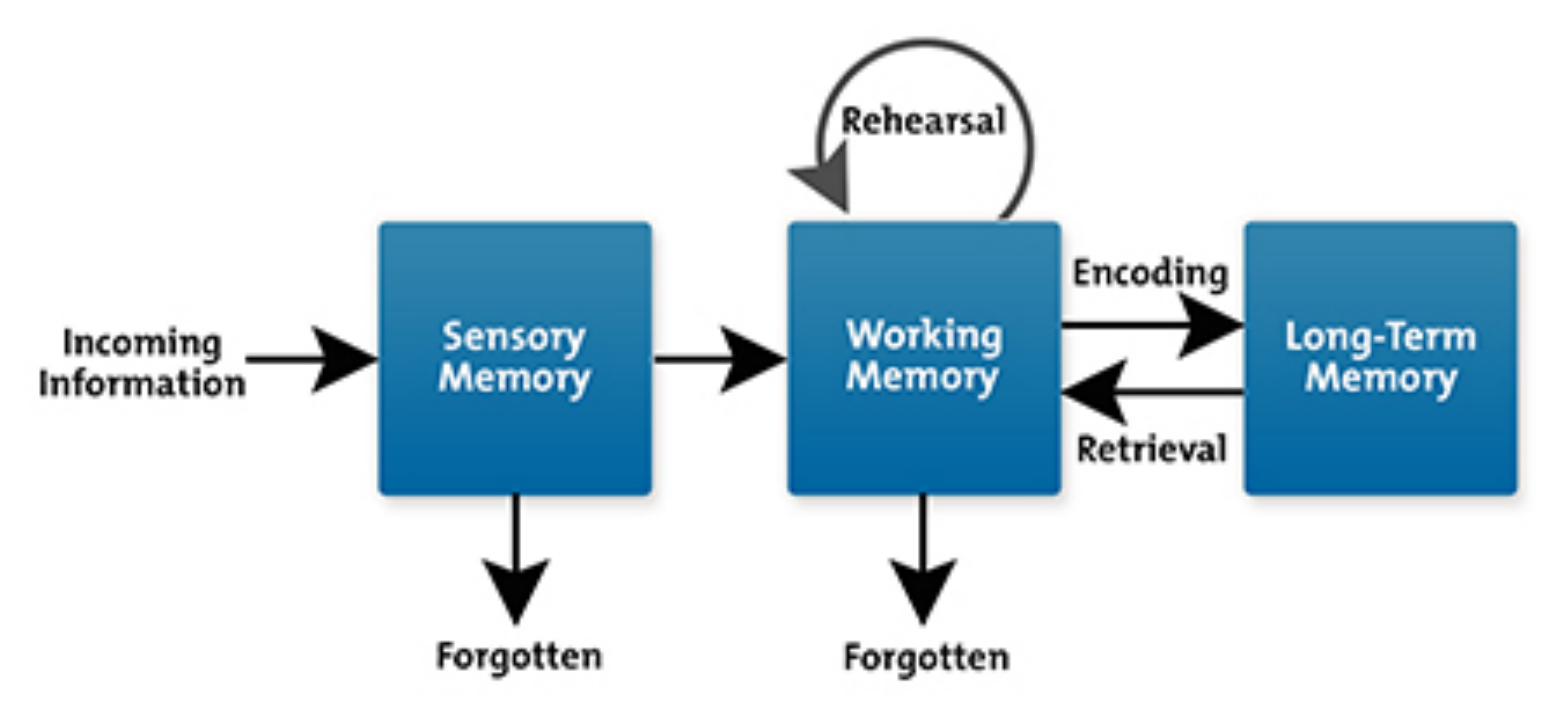}
\caption{Illustration of cognitive memory model}
\label{fig:memory}
\end{figure}

In CuedR, six keywords are randomly assigned to the user each from a distinct portfolio (e.g., animals, fruits, or vehicles), where each portfolio presents $26$ keywords. To aid memorability, our scheme offers graphical, verbal, and spatial cues corresponding to each keyword. In particular, each keyword (e.g., ``Zebra'') has an image (a picture of a zebra), a number and a phrase related to the keyword (``2. Each zebra has a unique pattern of stripes.''), and the position of both the image and phrase are fixed in both absolute terms and relative to the other images and phrases (Zebra is at the top between Cheetah and Elephant). See Figure~\ref{fig:portfolio} for a screenshot illustrating these features. 

Each time a portfolio is loaded, each of the 26 lowercase letters {\tt a-z} is assigned randomly as a {\em key} to one keyword on the page. The user inputs the key letter corresponding to her keyword into a single-character password field to move on to the next portfolio. The key letter changes every time to provide the {\em variant response} property~\cite{survey}. Schemes with this property have been shown to provide higher resilience to shoulder surfing and simple keystroke loggers than schemes like traditional textual passwords in which the same letters are entered at every login~\cite{shoulder06,survey}. 

\paragraphX{User authentication.} At user registration in CuedR, the system randomly selects six portfolios (without replacement) and the user is assigned one keyword from each of these portfolios. The user enters the key corresponding to the assigned keyword to get forwarded to the next portfolio. In case of a wrong entry she will immediately be informed about the error and will need to enter the correct one. During login, the user recognizes her system-assigned keyword from the portfolio and enters the key corresponding to that keyword into a small password field. A successful authentication requires the user to correctly enter keys for all six of her assigned keywords. 

When the user makes a mistake during login, CuedR shows the user a portfolio that is different from her next assigned portfolio. A legitimate user can recognize this {\em implicit feedback} as an indicator that something was wrong and that she should go back and correct the mistake, while an attacker will not know which portfolios are correct. Implicit feedback is thus a desirable feature to enhance usability when passwords have multiple parts~\cite{survey}. 

\paragraphX{Password storage.} For secure storage of the user's authentication secret, the six keywords can be concatenated together with a salt and hashed using a slow hash function like bcrypt~\cite{bcrypt} or PBKDF2~\cite{rfc2898}. Implicit feedback can be implemented by making the selection of the next portfolio a function of the current portfolio and the keyword selected, which is independent of the correctness of the responses. Thus, correctness only needs to be checked after all keys have been entered. 

\section{The Science behind C\lowercase{ued}R}\label{science}

In this section, we explain from the perspective of cognitive psychology how our design choices are set up to provide high memorability.

\subsection{Long-Term Memory}
We incorporate the scientific understanding of long-term memory to advance the scheme's usability properties. According to the cognitive memory model proposed by Atkinson and Shiffrin~\cite{long_mem68}, any new information is transferred to short-term memory (STM) through the sensory organs, where STM holds the information as {\em memory codes}, or mental representations of selected parts of the information. The information is transferred from STM to long-term memory (LTM), but only if it can be further processed and encoded. In this respect, an elaborative encoding would take place if the information can be associated with something meaningful, such as \textit{cues}. This encoding helps people to remember and retrieve the processed information efficiently over an extended period of time (see the illustration in Figure~\ref{fig:memory}).

In CuedR, users focus their attention to learn keywords through associating them with the corresponding cues, which should help to process and encode the keywords in memory and store them in the LTM. The cues would assist the user to recognize the keywords in the future, which should enhance their memorability. 

\subsection{Memory Retrieval} 
We designed CuedR to require users to perform a recognition task. Researchers in psychology have found that recognition (identifying the correct item among a set of distractors) is easier than recall (reproducing the item from memory)~\cite{tulving73} and have developed two main theories to explain this: {\em Generate-recognize theory}~\cite{grecog} and {\em Strength theory}~\cite{strength}.

Generate-recognize theory~\cite{grecog} speculates that recall is a two-phase process. In the generate phase, a list of candidate words is formed by searching long-term memory. Then, in the recognize phase, the list of words is evaluated to see if they can be recognized as the sought-out memory. According to this theory, recognition tasks do not utilize the generation phase and are thus faster and easier to perform. Strength theory~\cite{strength} states that although recall and recognition involve the same memory task, recognition requires a lower threshold of strength that makes it easier. The point is commonly illustrated in examples from everyday life. For example, multiple choice questions are frequently easier than essay questions since the correct answer is available for recognition. 

\subsection{Memory Cues} 
Psychology research~\cite{grecog,tulving73} has shown that it is difficult to remember information spontaneously without memory cues, and this suggests that authentication schemes should provide users with cues to aid memory retrieval. {\em Encoding specificity theory}~\cite{encoding} postulates that the most effective cues are those that are present at the time of remembering. In CuedR, cues are provided during registration, i.e., the learning period, and also at login.

\subsubsection{Why use multiple cues?} In CuedR, the user has five different references (cues) that she can leverage to learn the keyword: (i) the image, (ii) the number from 1 to 26, (iii) the phrase or fact associated with the keyword, (iv) the absolute positions of the keyword, the image, and the phrase/fact, and (v) the positions of the keyword, image and phrase/fact relative to the other keywords, images and phrases. This combination brings together graphical (images), spatial (positions), and verbal (facts, numbers) information. Thus, a user may focus on just those cues that she finds most appropriate to her learning process, while the other cues may provide additional support for memorability.

\paragraphX{Graphical cues.} Psychology research~\cite{dual_coding,nelson77} reveals that the human brain is better at memorizing graphical information as compared to textual information. This is known as the {\em picture superiority effect}, which motivates us to include graphical cues (images) in our scheme. Several explanations for this effect have been proposed. The most widely accepted is \textit{dual-coding theory}~\cite{dual_coding}, which postulates that in human memory, images are encoded not only visually and remembered as images, but they are also translated into a verbal form (as in a description) and remembered semantically. Another is the {\em sensory-semantic model}~\cite{nelson77}, which states that the images are accompanied by more distinct sensory codes that allow them to be more easily accessed. 

\paragraphX{Verbal and spatial cues.} While images are generally effective cues, not all users may have a strong visual memory. Additionally, many graphical password schemes require good vision and motor skills, which elderly users~\cite{elderly} may lack. Thus, we provide verbal and spatial cues in addition to graphical cues to let users leverage their cognitive ability in memorizing the keywords. Yan et. al.~\cite{mnemonic04} examined the influence of phrases in increasing the memorability of passwords, which inspires us to accommodate a common phrase or fact for each keyword as a verbal cue.

Having a fixed set of objects in a certain place aids to augment \textit{semantic priming}, which refers to recognizing an object through its relationship with other objects around it~\cite{passface}. Semantic priming thus eases the recognition task~\cite{passface}. In CuedR, the keywords and cues in a portfolio remain same and presented at a fixed position whenever that portfolio is loaded, which establishes a relationship between them and reinforces semantic priming. A recent study~\cite{graphic_soups13} also shows that keeping objects in a fixed position improves the usability during recognition.

\section{C\lowercase{ued}R: Through The Lens of Password Literature}\label{design}

Through a comprehensive survey on $25$ different graphical password schemes, Biddle et al.~\cite{survey} identified seven features that should be offered by an ideal graphical password system. The authors~\cite{survey} state, ``We expect tomorrow's ideal graphical password systems may have many of the following desirable characteristics, reflecting lessons learned from proposals to date." 

In this section, we analyze how CuedR addresses these seven features and explain our design choices based on the findings from the literature on passwords.

\subsection{[1] Theoretical password space meeting the security policy of the intended domain}

Well-known recognition-based schemes, such as Passfaces~\cite{passface} and Story~\cite{story}, originally provided no more than $13$ bits of theoretical entropy. Later, Hlywa et al.~\cite{image_type} conducted a study on recognition based graphical passwords with $20$ bits of entropy, since $20$ bits of entropy with reasonable lockout rules is considered sufficient to prevent online brute-force attacks~\cite{lockout}. In CuedR, we use more than $20$ bits of entropy, in particular $28$ bits, to maintain comparability with prior studies on system-assigned passwords~\cite{text_recog}. During login in CuedR, a user has to recognize her keyword from a portfolio of $26$ distinct keywords. This is required six times, once for each portfolio. The password space is thus $log_{2} {(26)}^6\approx 28$ bits. CuedR can be used with a range of entropy values by varying either the number of keywords or portfolios.

\subsection{[2] Avoiding exploitable reductions in security due to user choice of passwords}

Statistical password distributions are often not equiprobable due to scheme-dependent predictability of user choices~\cite{science_guessing}. In CuedR, passwords are randomly assigned by the system, which provides two security benefits. First, the effective password space in CuedR is same as the theoretical space. Second, the system gains user-choice resilience and thus provides robustness against online guessing attacks that exploit password reuse, personal information and predictable strategies~\cite{pwreuse14,pwpattern1}. 

\subsection{[3] At least mild resistance to shoulder surfing and key logging, through variant response}

Variant response refers to varying how the password is entered across different login sessions, which is an important feature to offer robustness against shoulder surfing and keystroke loggers~\cite{survey}. 

\subsubsection{Shoulder surfing} 
It is difficult in practice to observe both keyboard and monitor at the same time. Thus, graphical password schemes that include the variant response feature with keyboard entry provide higher resilience to shoulder surfing compared to traditional textual passwords and graphical passwords with mouse input~\cite{shoulder06}. In a study by Tari et al.~\cite{shoulder06}, participants playing the role of shoulder surfers were able to gain $73$\% of non-dictionary passwords, $26$\% of dictionary passwords, and $62$\% of graphical passwords with mouse input, but just $11$\% of graphical passwords with variant response.

CuedR offers the variant response feature, where the user enters a key corresponding to her keyword using the keyboard, and watching only keyboard entries is not sufficient for a shoulder surfer, as the key associated with each keyword changes with every login attempt. The entered key is shown as an asterisk or dot (as with a regular password) to minimize the risk of shoulder surfing. 

Variant response does not protect against an attacker who can use a video camera to record both the monitor and keystrokes at the same time, and attackers may gain the user's credentials when they are assigned during registration. Thus, we only claim that CuedR provides mild resistance to shoulder surfing through variant response, which conforms to the desired level of security in this regard~\cite{survey}. We recommend that users register in a secure environment (e.g., avoiding public terminals) to ensure better security against shoulder surfing. 

\subsubsection{Keystroke and mouse loggers} Keystroke loggers record keyboard input and mouse loggers capture mouse actions to make the user's credentials available for retrieval by remote attackers~\cite{survey}. Biddle et al.~\cite{survey} state that a system provides resilience against keystroke/mouse loggers when the keyboard/mouse entries for authentication vary across subsequent login sessions. Thus, the variant response feature in CuedR offers better resilience against basic keystroke loggers compared to a password system where the same letters are entered during every login session. CuedR is clearly resilient to mouse loggers, as it does not use mouse input.

\subsection{[4] Cues aiding memorability} While different variants of system-assigned passwords failed to provide satisfactory memorability~\cite{text_recog,PTP,forget_thesis,passphrase}, CuedR achieves a good memorability through associating each keyword with a set of cues and letting users choose the appropriate one(s) to their learning process~\footnote{We note that direct comparison between different studies should be taken with caution~\label{fn1}}. We describe the basis for the effectiveness of cues in the previous section, and we report on user perceptions of different cues in the results section.

\subsection{[5] Usability as close as possible to, or better than, textual passwords}

Shay et al.~\cite{longpw14} performed a comprehensive study on the memorability of user-chosen textual passwords following different composition policies, where \textit{basic12} was the simplest form of passwords in which the user had to create a password of at least $12$ characters without any composition requirements or dictionary check. Participants reported the least difficulty to create and remember a \textit{basic12} password. After two days, $86\%$ of participants who wrote down their password could log in ($76\%$ on the first attempt), while $75\%$ of participants who did not write down their password could log in ($61\%$ on the first attempt). For CuedR, after one week, we found that $100\%$ of participants could log in ($89\%$ on the first attempt). So, we see that memorability is better than textual passwords with moderate security requirements.

Although the login time for CuedR is high compared with traditional textual passwords, users mostly disagreed with the notion that the scheme is too time consuming (see Table~\ref{tab:usability}). Overall, users reported satisfaction with the usability of CuedR and $84$\% preferred to use it in real life as a replacement to traditional textual passwords.

\subsection{[6] Implicit feedback to legitimate users, when passwords are multi-part} 

Implicit feedback instantly notifies a user when she makes a mistake, instead of showing her an error message at the end of all entries. Due to its implicit nature, this feedback should only be recognizable and useful to the legitimate user. An attacker who does not know about a user's portfolios must make all six guesses in CuedR to learn whether he has succeeded or not. Implicit feedback has already been shown to have satisfactory user acceptance in a cued-recall based scheme~\cite{ccp}. To accommodate this feature in CuedR, we build distinct portfolios of images (i.e., ``animals", ``fruits", ``flowers", etc.) so that a user can clearly distinguish among the portfolios at a glance and quickly realize her mistake. 

\subsection{[7] Leveraging pre-existing user-specific knowledge where possible} 

Leveraging pre-existing user-specific knowledge, for example answering cognitive questions or recognizing personal images from decoys could make the scheme vulnerable to targeted guessing attacks (e.g., guessing by acquaintances). So, we did not include this feature in CuedR to ensure security. Since CuedR offers good memorability, it is not clear if user-specific knowledge is required, though it could help to reduce the cognitive burden on users. Exploring ways to securely leverage user-specific knowledge for authentication could be an interesting venue for future work.

\section{User Study}\label{study}

We now present the design of our user study to evaluate the usability and memorability of CuedR. The study procedures were approved by our university's Institutional Review Board (IRB) for human subjects research.

\subsection{Participants, Apparatus and Environment} 
For this experiment, we recruited $37$ students ($25$ women, $12$ men) through our university's Psychology Research Pool. Participants came from diverse backgrounds, including majors from Nursing, Psychology, Business, Political Science, Biology, Physical Science, and Social Work. The age of the participants varied between $17$ to $30$ with a mean age of $21$. They make regular use of the Internet and websites that require authentication. Each participant was compensated with course credit for participation and was aware that her performance or feedback in this study would not affect the amount of compensation.

The lab studies were conducted with one participant at a time to allow the researchers to observe the user's interaction with the system. For this study, we built $18$ different portfolios (e.g., animals, fruits, flowers, and vehicles), and collected the images (graphical cues) and phrases/facts (verbal cues) from free online resources. 

\subsection{Procedure} 
We conducted the experiment in two sessions, each lasting around $30$ minutes. The second session took place one week after the first one to test memorization of the password. Note that the one-week delay is larger than the maximum average interval for a user between her subsequent logins to any of her important accounts~\cite{pw_diary}. One week is also a common interval used in authentication studies (e.g.,~\cite{face_age,text_recog,bdas}).

\textit{Session 1.} After signing a consent form, the participants performed a practice trial with CuedR to compensate for novelty effect. We did not collect data for this practice trial. At registration, six portfolios were randomly chosen by the system and a user was assigned at most one keyword from each portfolio. Then participants were asked to spend $60$ seconds in completing a mental rotation test (MRT) puzzle shown on the computer screen, which helps to clear their working memory~\cite{mrt}. Participants were then given questionnaire that gathered demographic information, and were asked to log into the same site with CuedR ({\em login 1}). They were asked to not write down their authentication secrets.

\textit{Session 2.} The participants returned after one week of registration, and logged into the site using CuedR (\textit{login 2}). After they had finished, we conducted an anonymous paper-based survey. Participants were then compensated and thanked for their time.

\subsection{Ecological Validity}\label{ecology}  
Our participants were young and university educated, which represents a large number of frequent Web users, but may not generalize to the entire population. They came from diverse majors including Nursing, Psychology, Physical Science, Business, etc. As the study was performed in a lab setting, we were only able to gather data from $37$ participants. We believe that $37$ provides a suitable sample size for a lab study as compared to the prior studies on password memorability~\cite{geopass,pccp2,ccp,passpoint1}.

\section{Results}\label{results}

In this section, we discuss the results of our user study. We label the login performance of participants in session 1 and session 2 as \textit{login 1} and \textit{login 2}, respectively. We evaluated the usability of CuedR via all metrics suggested in the literature~\cite{design_space}: memorability, login time, number of login attempts, and user feedback. In addition, we analyzed the impact of portfolios on login performance and user perceptions on the effectiveness of different cues. We also discuss the results of pilot study on the memorability of multiple CuedR passwords.

\subsection{Memorability}\label{memory}

We observed a $100$\% login success rate for CuedR in both \textit{login 1} and \textit{login 2}. In \textit{login 1}, all the participants successfully recognized the keywords on the first attempt. In \textit{login 2}, $89$\% of participants succeeded on the first attempt to recognize all six keywords. The other four participants ($11$\%) recognized five of out of six keywords on the first attempt. Three participants corrected their mistake on the second attempt, and the other participant succeeded on the third attempt. 

\subsection{Registration and Login Time}\label{time} 

The mean time for registration was $31.2$ seconds (median: $30$ seconds, SD: $10.5$ seconds). The mean time for successful login were $25.7$ seconds (median: $24.0$ seconds, SD: $8.3$ seconds) in \textit{login 1}, and $38.0$ seconds (median: $39.0$ seconds, SD: $11.4$ seconds) in \textit{login 2}. A paired-samples t-test reveals that login time in \textit{login 1} was significantly less than that in \textit{login 2}, $t(36)=7.81$, $p<0.01$. This was expected, as participants performed {\em login 1} shortly after learning the keywords. To note, the reported registration and login time include the time to download images.

The login time in CuedR is in line with that in prior recognition based schemes offering $28$ bits of entropy~\cite{image_type,text_recog}. We note that our results for login time are likely conservative, since they measure initial use. A recent field study~\cite{geopass2} reveals that login time decreases with the frequent use of a scheme due to training effects. These findings are in agreement with our user feedback, where the participants reported that with practice, they could quickly recognize the keywords (see Table~\ref{tab:usability}). 

\begin{figure}[t]
\centering
\includegraphics[width=70mm, height=36mm]{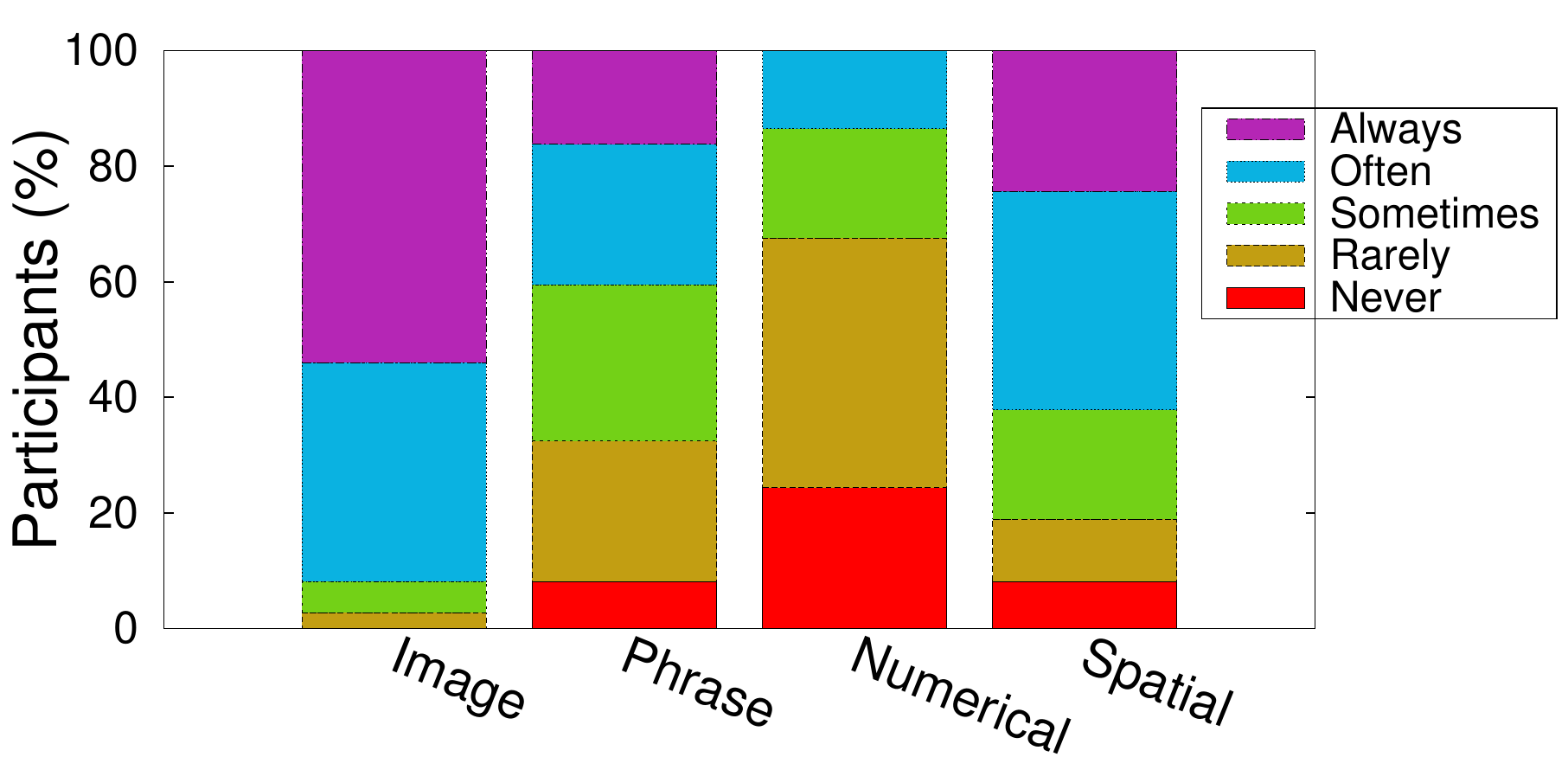}
\caption{Responses to the question: ``How often did the following cues assist you in recognizing keywords in CuedR?"}
\label{fig:cues}
\end{figure}

\begin{table*}[t]
\renewcommand{\arraystretch}{1.3}
\caption{Questionnaire responses for the usability of CuedR. Scores are
  out of 10. * indicates that scale was reversed. SD: Standard Deviation}  \centering
\vspace{0.2cm}
\begin{tabular}{c@{}m{8.5cm}rrrr@{}}
\hline
\hspace{0.2cm}&\multicolumn{1}{c}{Questions}&\multicolumn{1}{c}{Mode}&\multicolumn{1}{c}{Median}&\multicolumn{1}{c}{Mean}&\multicolumn{1}{c}{SD}\\ 
\cline{2-6}
& I could easily sign up with CuedR&$10$\phantom{a}&$9.0$\phantom{ab}&$9.0$\phantom{a}&$1.3$\phantom{m}\\
\cline{2-6}
& The login using CuedR was easy&$10$\phantom{a}&$10.0$\phantom{ab}&$9.5$\phantom{a}&$0.7$\phantom{m}\\ 
\cline{2-6}
& Keywords are easy to remember in CuedR &$10$\phantom{a}&$10.0$\phantom{ab}&$9.4$\phantom{a}&$0.8$\phantom{m}\\ 
\cline{2-6}
& *I found CuedR too time-consuming \newline (i.e., I did not find CuedR too time consuming) &$10$\phantom{a}&$7.0$\phantom{ab}&$6.4$\phantom{a}&$2.6$\phantom{m}\\
\cline{2-6}
& With practice, I could quickly enter my password in CuedR &$10$\phantom{a}&$10.0$\phantom{ab}&$9.8$\phantom{a}&$0.6$\phantom{m}\\
\cline{2-6}
& I could easily use CuedR every day&$10$\phantom{a}&$9.0$\phantom{ab}&$8.8$\phantom{a}&$1.3$\phantom{m}\\ 
\cline{2-6}
& I could easily use CuedR every week&$10$\phantom{a}&$9.0$\phantom{ab}&$9.0$\phantom{a}&$1.3$\phantom{m}\\ 
\hline
\end{tabular}
\label{tab:usability}
\end{table*}

\subsection{Impact of Portfolios on Usability}\label{im_portfolio}

In our study, all the participants succeeded to recognize their keywords irrespective of the type of portfolios in both \textit{login 1} and \textit{login 2}. In \textit{login 1}, no participant made any mistake in any portfolio, and thus there was no difference among portfolios for the number of attempts to succeed. In~\textit{login 2}, four participants ($11$\%) required multiple attempts to succeed (see the results for \textit{Memorability}), where one-way ANOVA test results show that there was no significant difference among portfolios in terms of the number of attempts required to successfully recognize the keywords, $F(17,220)=1.16$, $p=0.31$. In addition, we conducted a post-hoc pairwise comparison using Tukey's HSD (Honestly Significant Difference), which reveals no significant difference between any pair of portfolios for the number of attempts to succeed. 

Our one-way ANOVA test results demonstrate that there was no significant difference among different portfolios in terms of the time to learn the keyword during registration, $F(17,220)=0.76$, $p=0.71$, or recognize the keyword either in \textit{login 1}, $F(17,220)=1.16$, $p=0.31$, or in \textit{login 2}, $F(17,220)=0.59$, $p=0.87$. In addition, we conducted a post-hoc pairwise comparison using Tukey's HSD, which did not find any significant difference between any pair of portfolios in either registration time or in login time. These findings indicate that the usability in recognizing keywords did not vary significantly across different portfolios used in our study.

\subsection{User Perception on the Efficacy of Different Cues}\label{effect_cues} 

To understand user perception on the importance of different cues in aiding recognition, we asked them at the end of second session, ``How often did the following cues assist you in recognizing keywords in CuedR?" In response, for each cue they selected one of five options: \textit{Never}, \textit{Rarely}, \textit{Sometimes}, \textit{Often}, or \textit{Always}. Our results show that participants report using multiple cues to varying degrees to help recognize their keywords (see Figure~\ref{fig:cues}). In particular, $92$\% of participants reported that the images were \textit{always} or \textit{often} helpful to recognize keywords, while $62$\%, $40$\%, and $14$\% of participants, respectively reported that spatial, phrase, and numerical cues were \textit{always} or \textit{often} helpful in recognizing keywords. The participants' diverse choices for cues to aid recognition and their high login success rate support our anticipation that letting users choose the appropriate cue(s) to their learning process aids the memorability for system assigned random passwords. 

\subsection{User Feedback on Usability and Applicability}\label{feedback}

We asked the participants to answer two sets of $10$-point Likert-scale questions ($1$: \textit{strong disagreement}, $10$: \textit{strong agreement}) at the end of the second session. We reversed some of the questions to avoid bias; the scores marked with (*) were reversed before calculating the modes, medians, and means. So, a higher score always indicates a more positive result for CuedR. To design the questionnaire, we carefully followed the guidelines provided in the existing password literature~\cite{geopass,ccp,passpoint3}, including using nearly identical questions to those from other studies.

\begin{table}[b]
\renewcommand{\arraystretch}{1.3}
\caption{The applicability of CuedR for different online
  accounts. Scores are out of 10.}
\vspace{0.2cm}
\centering
\begin{tabular}{c@{}lcccc}
\hline
\hspace{0.2cm}&\multicolumn{1}{c}{Online accounts}&Mode&Median&Mean&SD\\ 
\hline
& Bank&$10$&$8.0$&$7.4$&$2.6$\\ 
\cline{2-6}
& E-mail&$10$&$9.0$&$8.1$&$2.1$\\ 
\cline{2-6}
& Social Networking&$10$&$9.0$&$7.7$&$2.4$\\ 
\cline{2-6}
& University Portal&$10$&$8.0$&$8.2$&$1.9$\\ 
\cline{2-6}
& E-commerce&$10$&$9.0$&$7.8$&$2.5$\\ 
\hline
\end{tabular}
\label{tab:applicable}
\end{table}

\subsubsection{Usability} Participants showed a high degree of satisfaction with the usability (e.g., memorability, ease of login, ease of using either weekly or daily) of CuedR. Their feedback was also positive (mode, median, and mean higher than neutral) regarding login time, and they indicated that with practice they could log in quickly using CuedR (see~Table~\ref{tab:usability}). In our study, we could not test the usability of implicit feedback for CuedR, since most users did not make enough login mistakes to gain experience with it.

\subsubsection{Applicability} At the end of second session, we asked $31$ of the participants,\footnote{We failed to ask the first six participants.} ``Do you want to use CuedR in real life as a replacement to traditional textual passwords?" $84$\% responded `Yes', $10$\% responded `Maybe', and two participants responded `No', where both of them mentioned that they would prefer traditional textual passwords in real life as they did not find any problems with them. User feedback about the applicability of CuedR in different online accounts is illustrated in Table~\ref{tab:applicable}.

\subsection{Pilot study: Memorability for Multiple CuedR Passwords}
It is common in password research to report a single-password study in the first article of a new authentication scheme, which helps to establish performance bounds and figure out whether multiple-passwords tests are worthwhile in future research. A recent survey~\cite{survey} reported that out of $25$ graphical password schemes proposed to date, only three have been evaluated through a multiple-password study, and none of these study results was reported in the first article. Since the use of multiple passwords is an important issue for deployment, however, we conducted a pilot study for multiple passwords, in addition to reporting the detailed results of a single-password study.

The study procedure was same as that in our single-password study, except that each participant was assigned three CuedR passwords ($18$ keywords, in total) instead of one. To administer this experiment, we created three different websites outfitted with CuedR and presented the sites to participants as tabs in an open browser window. Participants were free to select the order of websites at registration and login, but the tabs were arranged the same way every time. For this study, we recruited $11$ students ($9$ men, $2$ women) who came from various majors of our university. We believe that $11$ represents a suitable sample size for a pilot study~\cite{pccp_pilot}. 

In this study, all of the participants were able to log in successfully within three attempts in both \textit{login 1} (same day of registration) and in \textit{login 2} (one week after registration). In \textit{login 1}, nine participants ($82$\%) succeeded on the first attempt for all three CuedR passwords. One participant ($9$\%) succeeded to log in using two CuedR passwords on the first attempt, where she recognized $17$ keywords on the first attempt and corrected the lone mistake on the second attempt. Another participant succeeded on the first attempt for one CuedR password, where she successfully recognized $15$ keywords on the first attempt and corrected the mistakes on the second attempt.

In \textit{login 2}, six participants ($55$\%) succeeded on the first attempt to recognize all $18$ keywords. Four participants ($36$\%) successfully recognized $17$ keywords on the first attempt, i.e., they succeeded to log in using two CuedR passwords on the first attempt. For another CuedR password, two ($18$\%) of these four participants succeeded on the second attempt and other two participants succeeded on the third attempt. One participant ($9$\%) successfully recognized $16$ keywords on the first attempt. In particular, she succeeded to log in using one CuedR password on the first attempt and succeeded on the second attempt for other two CuedR passwords.

\section{Discussion}\label{disc}

In this section, we discuss three important aspects of CuedR: i) impact, ii) acceptance, and iii) application. Here, the term \textit{study} refers to our single-password study, unless otherwise specified. We conclude with a discussion on the scope for future research on CuedR. 

\subsection{C\lowercase{ued}R: The Impact}
Existing password systems fail to fully address users' cognitive limitations or leverage humans' cognitive strengths. Thus, despite a large body of research, it still remains a critical challenge to build an authentication scheme that provides both guessing resilience and high memorability. CuedR represents a breakthrough, offering high memorability for system-assigned random passwords, and shows a promising research direction to leverage humans' cognitive abilities for user authentication. 

System-assigned passwords provide higher security against guessing attacks than user-chosen passwords, but it is difficult for most people to memorize them~\cite{passphrase,text_recog,interference3}. Users have varying cognitive strengths and abilities, and it is hard to know in advance what will help a given user to remember her password. In CuedR, we present a variety of visual, verbal, and spatial information related to randomly selected keywords in an organized way and then let users choose the appropriate cue(s) to their learning process. Further, the cues can work together. When we asked users to identify the cues they used, $83.8$\% of users reported of using multiple cues often or always. As one participant commented, ``The image, phrase and number naturally correspond in my mind, and make it easy to remember.''

CuedR also shows that the cued-recognition class of password schemes, a new design point in the field, can be effective for user authentication. In particular, CuedR addresses each of the features needed in an effective graphical password scheme, as identified by Biddle et al.~\cite{survey} from their comprehensive survey on the graphical password literature.

Here we mention a participant's feedback that particularly drew our attention: \textit{``The multiple cues make it a helpful password scheme for autistic persons, who find it cognitively difficult to create secure passwords."} We appreciate such a thoughtful opinion and note it to be an important issue to be explored in future work.

\subsection{C\lowercase{ued}R: The Acceptance}
In traditional user-chosen passwords, users bear the responsibility of ensuring security for their online account through a secure password that should be chosen with creativity and intelligence so that it achieves satisfactory memorability. For many users, this is a lot of work, and thus in many cases they compromise with security and create a weak but memorable password. A recent study~\cite{hijack14} reveals that with the advancement of digital technology and widespread use of internet in recent years, users now better realize the importance of strong passwords than anytime before, and many of them intend to create secure passwords but just fail to achieve a good balance between security and memorability. So, rather than blaming users for predictable passwords, researchers should improve how authentication systems address human cognitive abilities. 

Participants in our study seem to be convinced with the security provided by a system-assigned password. In a post-experiment open-ended question where they were asked about their opinion of CuedR, most of them reported high satisfaction with its security features. 

Since the participants were convinced with the security of six system-assigned keywords, and could efficiently recognize each keyword in a reasonable time ($6.3$ seconds, on average) after a week of registration, they found the overall login time acceptable and reported satisfaction with the usability of CuedR (see Table~\ref{tab:usability}). $84$\% of participants preferred to use the scheme in real life as a replacement to traditional textual passwords. 

\subsection{C\lowercase{ued}R: The Applications}
Although most of the participants reported strong agreement about using CuedR for all of the given account types (see Table~\ref{tab:applicable}), we must be cautious to recommend its application since textual passwords have lower login times than CuedR. Traditional textual passwords are fraught with security problems that make them less than desirable, especially for high-security accounts~\cite{pwreuse14}. Ideally, there should be a clear separation between the passwords used for low-security websites and high-security websites~\cite{pw_thicket}. Thus, CuedR can be used as a standard authentication mechanism for online accounts with high security requirements and where logins occur relatively infrequently, such as financial (e.g., online banking, brokerage services) and e-commerce accounts~\cite{pw_diary}. A study by Hayashi and Hong~\cite{pw_diary} finds that users log into financial and e-commerce sites once a week on average, which is in agreement with the interval of one week before \textit{login 2} in our study. By using CuedR for high-security accounts, it helps to build the mental separation with lower security accounts and avoid attacks based on password reuse and predictable patterns. 

As compared to other cognometric graphical password schemes that present users with images only, the deployment of CuedR may require more effort, where separate portfolios of keywords are built accommodating both graphical and verbal cues. We note that each commercial deployment can use a small set of portfolios
for all of its users. For example, with $10$ portfolios, a phisher could correctly guess the first two portfolios for a user only $1$\% of the time. 

\subsection{Future Work}

Now that lab-study results show promise for CuedR, it would be an interesting avenue for future work to evaluate the scheme through a long-term multiple-password study with larger and more diverse populations, where we would explore the training effect in reducing login time over more login sessions. A recent field study~\cite{geopass2} reveals that login time decreases with the frequent use of a scheme due to training effects.

In the current interface of CuedR, users have to look at a separate table to find the phrase/fact related to a keyword (see Figure~\ref{fig:portfolio}). In future work, we will test an alternate interface design to improve login time: The fact related to a keyword would be shown just below the graphical cue of that keyword, in which case users should require less time to find a phrase than finding it from a separate table.

While we have found that combining multiple cues shows promising results for authentication, we plan to pursue future studies to address the following issues: i) The impact of cues on the login performance of users from different age groups; ii) The usability of offering various combinations of cues; iii) The correlations between usability and the elimination of cue(s) from the interface over login sessions; iv) The usability of leveraging different cues from mobile devices. 

\section{Conclusion}\label{conc}

In this paper, we present a novel authentication scheme, CuedR, which helps us to explore the efficacy of combining graphical, verbal, and spatial cues to improve the memorability of system-assigned random passwords. We also discuss the promise of CuedR in addressing the features of an effective graphical password scheme~\cite{survey}. Although the login time is relatively high, our primary findings indicate high memorability for CuedR, suggesting that cued-recognition would be an important direction in password research to address the usability-security tension in authentication. 

\balance


\section{Acknowledgement}
This material is based upon work supported by the National Science Foundation under Grant No. CNS-1117866 and CAREER Grant No. CNS-0954133. We are thankful to the anonymous reviewers for their thoughtful suggestions in improving the paper.

\bibliographystyle{acm-sigchi}  
\bibliography{refs}
\end{document}